\documentstyle[12pt,aaspp4,epsf]{article}

\begin{document}

\def\intldate{\number\day\space\ifcase\month\or
January\or February\or March\or April\or May\or June\or
July\or August\or September\or October\or November\or December\fi
\space\number\year}

\def \deg    {$^{\circ}$}
\def \eg     {{e.g., }}
\def \cf     {{cf.\ }}
\def \ie     {{i.e., }}
\def \hub    {{$H_{\hbox{\eightrm 0}}$}}
\def \hunits {{km s$^{\hbox{\eightrm --1}}$ Mpc$^{\hbox{\eightrm --1}}$}}
\def \sec    {$^{s}$}
\def \arcsecpoint {$.^{''}$}

\def \hi    {\ion{H}{1}}
\def \hii    {\ion{H}{2}}
\def\tightenlines{\def\baselinestretch{1}\small\normalsize}
\def\gtorder	{\mathrel{\raise.3ex\hbox{$>$}\mkern-14mu\lower0.6ex\hbox{$\sim$}}}
\def\ltorder	{\mathrel{\raise.3ex\hbox{$<$}\mkern-14mu\lower0.6ex\hbox{$\sim$}}}
\def\sb		{{\rm mag~arcsec$^{-2}$}}
\def\area	{${\rm deg}^2$}
\def\kpc	{\hbox{\rm kpc}}
\def\kms	{\hbox{$\rm km \, s^{-1}$}}
\def\pc		{\hbox{ pc }}
\def\yr		{ \, {\rm yr}}
\def\peryr	{ \, {\rm yr^{-1} }}
\def\vlos	{ v_{\rm los} }
\def\lsim	{\rlap{\lower .5ex \hbox{$\, \sim \, $} }{\raise .4ex \hbox{$\, < \, $} }}
\def\gsim	{\rlap{\lower .5ex \hbox{$\, \sim \, $} }{\raise .4ex \hbox{$\, > \, $} }}
\def\solar	{ {\odot} }
\def\lsolar	{ {\rm L_{\odot}} }
\def\msolar	{ \rm {M_{\odot}} }
\def\rsolar	{ {R_{\odot}} }
\def\rnot		{ {R_{o}} }
\def\vsolar	{ {v_{\odot}} }
\def\vnot	{ {v_{o}} }
\def\surfmunit  { \rm {\, \msolar \, pc^{-2}} }
\def\HI		{{H{\sc I}}}
\def\etal	{{et~al.}}
\def\mags	{{ \, \rm mag }}   
\def\percubicpc	{ { \pc^{-3} } }
\def\abs	{ \hbox{ \vrule height .8em depth .4em width .6pt } \,} 
%
%

%
%

\title{Does the Milky Way have a Maximal Disk?}

\author{Penny D.~Sackett}
\affil{Institute for Advanced Study, Princeton, New Jersey 08540, USA\\
and\\
Kapteyn Astronomical Institute, 9700 AV Groningen, The Netherlands\\
psackett@astro.rug.nl}

\begin{abstract} 

The Milky Way is often considered to be the best example of a 
spiral for which the dark matter not only dominates the outer kinematics, 
but also plays a major dynamical role in the inner galaxy:  
the Galactic disk is therefore said to be ``sub-maximal.''  
This conclusion is important to the understanding of 
the evolution of galaxies and the viability of particular dark matter models. 
The Galactic evidence rests on a number of 
structural and kinematic measurements, many of which have 
recently been revised.  
The new constraints indicate not only that the Galaxy is a more 
typical member of its class (Sb-Sc spirals) than previously thought, 
but also require a re-examination of the question of whether or not
the Milky Way disk is maximal. 
By applying to the Milky Way the same definition of ``maximal disk'' 
that is applied to external galaxies, 
it is shown that the new observational constraints are  
consistent with a Galactic maximal disk of reasonable $M/L$.  
In particular, the local disk column can be substantially less than 
the oft-quoted required $\Sigma_{\odot} \approx 100 \, \msolar pc^{-2}$ 
--- as low as $40 \, \msolar pc^{-2}$ in the extreme case --- 
and still be maximal, in the sense that the dark halo provides 
negligible rotation support in the inner Galaxy. 
This result has possible implications for any conclusion that rests 
on assumptions about the potentials of the Galactic disk or dark halo, 
and in particular for the interpretation of microlensing 
results along both LMC and bulge lines of sight.  

\end{abstract}


\vskip 1cm

\begin{center}
{\it Submitted to the Astrophysical Journal on 25 August 1996}\\
{\it Accepted for publication on 20 January 1997}
\end{center}

\newpage


\section{Introduction}

The total amount and distribution of mass in disk galaxies 
determines their kinematics, and thus the amplitude and 
shape of their stellar and gaseous rotation curves. 
Since rotation curves of spirals fall much less rapidly than would be  
expected from the distribution of their light, it is generally 
inferred (assuming Newtonian gravity) that unseen, extended 
mass is present that dominates the 
kinematics in the outer regions of the disk.  
For both external galaxies and the Milky Way, however, the total 
amount of this ``dark'' mass within a given radius is ill-constrained, 
partially due to uncertainty in the geometry of the dark mass, and 
partially due to uncertainty in the fraction of the total mass supplied 
by the stellar disk itself (\cf Sackett 1996).  
Loosely speaking, if the 
disk is thought to dominate the mass in the inner regions of the 
galaxy, it is said to be ``maximal,''  
whereas if the dark halo is dynamically important in the inner galaxy 
the disk is said to be ``sub-maximal.''

It is important to settle whether spiral disks are maximal 
because the disk mass is coupled directly (through the rotation curve) 
to the radial distribution of halo dark mass, 
the form of which is predicted by models for galaxy formation and 
the nature of dark matter.   
In general, models of dissipationless dark matter predict 
that spiral disks should not be maximal: the dark mass in these 
models is substantially less flattened vertically 
than the disk baryons and has a small core radii so that it dominates 
even the inner galactic kinematics.  
For external spirals, evidence has been presented both for 
and against maximal disks (Casertano \& van~Albada 1990;  
Freeman 1993; van~der~Kruit 1995), with most of 
the controversy centered on the uncertainty 
in the stellar disk mass (or more correctly, the total mass 
{\it associated\/} with the stellar disk).  
Since the local disk mass can be 
more directly measured in the Milky Way, the question has 
been less controversial for the Galaxy:  
it is generally believed that the Milky Way disk is substantially 
sub-maximal (Bahcall 1984; van der Kruit 1989; Kuijken \& Gilmore 1991, 
Merrifield 1993, Kuijken 1995).    

Here we re-examine the evidence that has led to the suggestion that the 
Milky Way disk is not maximal, in light of new observations that 
could  alter considerably the estimated contribution of the disk 
mass to the inner rotation curve of the Galaxy.   Since the 
discussion of this point has sometimes been hindered by imprecision 
in the definition of ``maximal,'' we define and apply a 
consistent definition of the term to both the Galaxy and 
external spirals.  We will show that new observations of the 
structural parameters of the Milky Way disk suggest that the Galaxy is  
consistent with the structure inferred for external disks, and in particular 
with a ``maximal'' disk.  
Our approach differs from Sellwood \& Sanders (1988) in that we do 
not attempt to fit the uncertain Galactic rotation curve with a single 
disk model, but rather explore a variety of models 
consistent with the newer constraints.  
A definition of maximal disk, as it is applied in external galaxies, is 
formulated in \S 2.  In \S 3, recent observational determinations of the 
kinematic and structural parameters of the Galaxy are reviewed.  
In light of these new Galactic constraints, the maximum disk hypothesis 
is tested in the Milky Way in \S4 and shown to be viable.  Concluding 
remarks can be found in \S5. 

\section{What is a Maximal Disk?}

Stellar disks, including that of the Milky Way, are generally modeled 
as double exponential disks, with volume densities given by  
\begin{equation}
\rho (R,z) = \rho_o \, e^{- R/h_R} \, e^{- \left| z \right| /h_z} ~~~~,
\end{equation} 
\noindent where $R$ and $z$ are the natural cylindrical coordinates 
of the axisymmetric disk, and $h_R$ and $h_z$ are 
the scale length and scale height of the disk, respectively.  
The integral column of disk mass (integrated to infinity) for such 
a disk at any galactocentric radius $R$ is 
$\Sigma(r) = \Sigma_o \, e^{- R/h_R}$, where 
$\Sigma_o = 2 \, \rho_o \, h_z$ is the central (face-on) 
surface mass density.  
The total disk mass is then
simply $M = 2 \, \pi \, \Sigma_o \, h_R^2$.

The equatorial rotation curve due to this massive disk can be found 
by solving Poisson's equation for the corresponding gravitational 
potential $\Phi$, and then differentiating $\Phi$ to find the 
radial force giving rise to the circular speed at $R$.   
In this way, a general 2-D integral for the radial force of thick disks was 
computed by Casertano (1983); Kuijken \& Gilmore (1989a) reduced 
this to a 1-D integral for double exponential disks.  
With the assumption of circular orbits, the radial force leads \
directly to the rotation curve via Newton's second law.  
Although the rotation curve of a substantially thickened exponential disk 
peaks at slightly larger radius than that of an infinitely thin one, 
a thick disk curve is also less sharply peaked.  The net result is that 
the speed of the disk at $R = 2.2 \, h_R$ is equal to the peak rotation speed 
of the disk to $<1\%$ over a wide range of scale heights, $h_z$. 
We will thus refer to $v^{disk}(2.2h_R)$ as the maximum rotation speed 
of the disk.

The total circular speed of material at any point in a spiral 
galaxy is a result of the combined rotation support at that radius of the 
massive disk, halo, and bulge components.  
(Here ``halo'' is used to indicate dark halo; 
the observed metal-weak stellar halo contains so little mass as to be 
dynamically negligible.)  
Thus, in order to determine the radial distribution of halo dark mass 
from the rotation curve of a galaxy, the disk and bulge contributions 
must be well-understood and properly subtracted.   
In external galaxies, the structural parameters of the bulge and disk 
light can be measured directly via surface photometry.  The conversion of 
these measurements into structural parameters for bulge and disk 
{\it mass\/}, however, is not uniquely determined.  
It is generally assumed that the mass-to-light ratios ($M/L$) 
of the bulge and disk components do not vary with radius, in which case   
the luminous structural parameters of the bulge and disk  
control the shape of the bulge and disk rotation curves, 
with the single number, $M/L$, controlling the amplitude.  

The ``maximal disk hypothesis'' is commonly stated as 
the hypothesis that the luminous disk {\it alone\/} is responsible 
for the circular velocity in the inner region of the rotation curve 
(van Albada \& Sancisi 1986).  
Strictly speaking, however, this definition 
is never applied to external galaxies for three reasons: 
(1) The luminous disk mass cannot be separated from dark disk mass with 
	the same distribution (\ie with the same $h_R$).  
(2) In galaxies with prominent bulges, the luminous bulge mass also 
	contributes to the inner rotation support.   
(3) Since non-hollow halos are generally used for rotation curve fits, 
	a small amount of inner rotation support must arise 
	from the core of the halo itself.

In practice, then, the maximum disk hypothesis is imposed by 
assuming a functional form for the dark halo 
(isothermal with core is a common choice), and then fitting the 
entire rotation curve while adjusting the free halo parameters 
so as to accomodate the largest possible disk $M/L$.  (If the galaxy 
has a prominent bulge, the bulge $M/L$ may be maximized first, or 
for simplicity set equal to the disk $M/L$.)
This leads to a ``maximum'' disk mass that provides somewhat 
less than the total rotation support at 2.2$h_R$ where the disk 
rotation curve peaks, and substantially less support at larger radii 
where the dark halo is presumed to dominate.  

For external galaxies with type similar to that of the Milky Way (Sb to Sc), 
such maximum disk fits produce massive disks whose peak disk rotation speed 
$v^{disk}_{\rm max} = v^{disk}(2.2h_R)$,   
is 75\% to 95\% of the {\it total\/} circular speed at 
that radius (\cf Begeman 1987; Broeils 1992).  
The lower end of this range is occupied by galaxies with large 
bulges or bars that contribute significantly to $v$(2.2$h_R$), 
thus decreasing the importance of the disk in the inner regions. 
In general, the total circular speed at $R = $2.2$h_R$ may be 
somewhat different than the peak rotation speed of the galaxy as a whole, 
depending on whether the rotation curve falls slightly (typical for 
massive bulge-dominated spirals) or rises (typical for extreme late-type 
spirals), but for Sb-Sc spirals the two are comparable.  
In very late-type or dwarf spirals, a significant fraction of the 
rotation support may be provided by the mass of the neutral gas, but 
this is negligible in the inner portion of Sb-Sc spirals. 

In summary, the definition of maximum disk that is 
typically used in external spirals does not necessarily require 
that the detected luminous disk matter provide all the rotation 
support in the inner galaxy.  If spirals contain substantial 
disk dark mass with a scale length similar to that of the light, the 
disk rotation curve shape will remain unchanged, but the inferred 
$M/L$ ratio of the massive disk will be larger than the integrated 
stellar $M/L$ of the disk.  Secondly, a maximum disk fitted to the 
rotation curve of an external spiral with a bulge component may supply 
as little as 75\% of $v$(2.2$h_R$) and thus contain only slightly more 
than one-half of the mass interior to that radius. 
The practical application of the maximum disk hypothesis 
to spirals similar to the Milky Way 
produces disks that provide 85\% $\pm$ 10\% of the total rotation support 
of the galaxy at $R = 2.2 h_R$.  
Rather than attempt a multi-component fit to the uncertain 
Galactic rotation curve, this extragalactic working definition of 
maximal disk will be applied directly to current observations of the 
Milky Way. 

\section{Observational Constraints on the Structure of the Galactic Disk}

In external galaxies, the rotation curve and 
scale length of the disk light are generally well-determined, 
but the mass normalization of the disk is poorly constrained. 
In the Milky Way, the situation is reversed.   
The local surface mass density of the Galaxy is well-constrained by 
kinematical studies of old stars in the solar neighborhood, although 
parceling this mass column into halo and 
disk components has proven less straightforward.  
As in external galaxies, it is generally assumed that the luminous 
and massive disk have the same scale length, but in the Milky Way 
determinations 
of the luminous scale length are hindered by our observing vantage 
point, which places us in the middle of the dusty disk.  
The rotation support supplied by the massive disk must be compared 
to the observed rotation curve for the Galaxy, but here too 
our position inside the disk introduces uncertainties 
and ambiguities that plague attempts to construct an accurate 
rotation curve.   Finally, in order to 
convert measurements of the circular rotation referenced to the Local 
Standard of Rest (LSR) and local surface mass density 
into an absolute rotation curve and total disk mass, 
an accurate measurement of our distance 
from the Galactic Center is required.  

In the following sections, new determinations of important structural 
and kinematic properties of Milky Way will be 
discussed in turn:  local surface mass density, $\Sigma_{\odot}$; 
scale parameters of the luminous disk, $h_R$ and $h_z$; 
the distance to the center of the Galaxy, $\rnot$; the 
rotation curve, $v(R)$; and local circular speed, $\vnot$.  
As we shall see, 
these crucial measurements have evolved over the past 
few years in a direction that requires a reassessment of whether 
the Galactic disk is maximal.  
A short note on the bulge of the Milky Way and its effect on this 
analysis ends this section. 

\subsection{Local Surface Mass Density}

The distances and kinematics of old, resolved stars 
can be used to measure the vertical restoring force 
of the local Galactic disk --- and thus its surface mass density.  
By measuring the velocities of local stars as a function of 
height above the Galactic plane, Kuijken \& Gilmore (1991, KG) 
report a total mass column of 
$\Sigma_{\rm 1.1~kpc} = 71 \pm 6 \, \msolar pc^{-2}$, 
integrated within a 1.1~kpc band above and below the plane.  
This is a rather robust measurement that depends only weakly 
on assumed scale parameters for the disk.  The separation 
of this surface mass density into disk and halo components, however, 
is much more dependent on the assumed scale length $h_R$ and 
scale height $h_z$ of the disk, and on the distance $\rnot$ 
to the Galactic center.  
Assuming $h_R$ = 4.5 \kpc, $h_z$ = 0.3~kpc, and $\rnot$ = 7.8 \kpc,  
KG concluded that $\Sigma_{\odot} = 48 \pm 9  \, \msolar pc^{-2}$ 
was due to the disk itself, with the rest contributed by a rounder dark halo.  
Gould's reanalysis (1990) of the original Kuijken \& Gilmore dataset (1989b) 
weighs in at a similar $\Sigma_{\odot} = 54 \pm 8 \, \msolar pc^{-2}$.

If the local dark matter contains a significant disk-like component, 
these numbers would increase.  
Using a different dataset than that of KG, and 
fitting models in which the local dark 
matter distribution is proportional to the disk luminous matter 
(assuming $h_R$ = 3.5 \kpc, $h_z$ = 0.35~kpc, and $\rnot$ = 8 \kpc) 
Bahcall (1984) suggested that the total local disk column may be 
as much as $67 \, \msolar pc^{-2}$.  
Applying the same models to newer data, 
Bahcall, Flynn \& Gould (1992) derived an even 
larger total disk column at the Sun 
of $84^{+29}_{-24} \, \msolar pc^{-2}$, although a recent 
redetermination of the normalization of the local density of 
the kinematical tracer population later led 
Flynn \& Fuchs (1994) to revise this estimate downwards to 
$56 \pm 10  \, \msolar pc^{-2}$.  
If the dark matter is disk-like but with a scale height larger than 
that of the luminous matter, the total disk column may be somewhat 
higher than these estimates. 
A strict lower limit on the local surface mass density is given 
by the total column of {\it directly observed\/} material in the solar 
neighborhood.  
The lowest recent estimate is that of Gould, Bahcall \& Flynn (1996) who 
report that the column of M dwarfs inferred from 
their deep HST observations is smaller than previously thought, 
yielding a total local {\it observed\/} column 
of $\Sigma_{\solar, obs} = 39 \, \msolar pc^{-2}$.

Independent recent measurements thus seem to confirm that the 
local surface mass density of the thin disk is about 
$\Sigma_{\odot} = 53 \pm 13  \, \msolar pc^{-2}$, with a weak 
dependence on the assumed structural parameters of the luminous disk 
and dark matter distribution.  
As stressed by KG, the more robust estimate  
is $\Sigma_{\rm 1.1 \, kpc} = 71 \pm 6 \, \msolar pc^{-2}$ for 
the total local column between 1.1~kpc above and below the plane.

\subsection{Scale Lengths and Heights of the Thin and Thick Disks}

Disk scale lengths $h_R$ of 3.5 and 4.5~kpc have been used recently 
by authors studying the disk mass distribution 
(\cf Bahcall, Flynn \& Gould 1992; Kuijken \& Gilmore 1989b, 1991).  
The newer observations which we now discuss, however, 
suggest that the scale length of the 
disk stars in the Milky Way may be considerably shorter.  
Since measurements of $h_R$ scale with the solar distance $\rnot$, 
the assumed $\rnot$ is listed in Table 1 for each determination of $\rnot$. 
As we will see in \S4, the dynamical importance of the disk is sensitive 
to the ratio $\rnot/h_R$, which is more observationally secure than $h_R$. 

The three-dimensional distribution of 2.4 micron light from 
SpaceLab was used by Kent, Dame \& Fazio (1991) to derive a scale 
length for the Galactic disk of $h_R = 3.0 \pm 0.5~$kpc.  
They derive a scale height of IR light $h_z = 0.247~$kpc at the solar position, 
decreasing to $h_z = 0.165~$kpc at a distance of 5~kpc from the 
Galactic center.  
Kent, Dame \& Fazio review earlier work on the disk scale length $h_R$ 
and note that values as small as 1.8~kpc and as large as 
6.0~kpc have been suggested.  Infrared estimates based on integrated 
light or star counts tend to give shorter scale lengths than their 
optical counterparts.  
This mirrors the situation in external galaxies in which longer 
scale lengths are found in bluer bands, perhaps indicative of 
radial gradients in age and metallicity (\cf de~Jong 1996).   
Measurements in redder bands have the advantage that they are more likely 
to reflect the distribution of stellar mass (since they are dominated 
by the emission from low-mass stars that in turn dominate the stellar 
mass) and are less contaminated by extinction from dust, especially 
important in edge-on systems (like the Milky Way).

A review of early determinations of Galactic structure constants based on 
optical star count studies prior to 1990 can be found in 
Robin, Cr\'ez\'e, \& Mohan (1992).  
Although these earlier optical studies supported moderate 
scale lengths of 3.5 --- 4.5~kpc (as does the single-field study of  
Ng \etal\ 1995), Robin, Cr\'ez\'e, \& Mohan found evidence for much shorter 
scale length of $h^{thin}_R = 2.5 \pm 0.3~$kpc and a sharper 
decrease in counts beyond 6 kpc. 
Most of the more modern determinations of $h_R$ from optical 
star count studies are smaller than the pre-1990 optical estimates and 
closer to IR-estimated scale lengths.   
By modeling local kinematic data using an approximation that the 
stellar disk is cold and self-gravitating, Fux \& Martinet (1994)  
concluded that the old disk has a scale length of 
$2.5 ^{+0.8}_{-0.6}~$kpc, under the assumption of a constant 
scale height.  If they instead assumed an inwardly decreasing scale 
height, the best-fitting scale length increased to 3.1~kpc.
Using UBVR photometry and proper motions from three fields in the
direction of the Galactic center, anti-center, and anti-rotation, 
combined with previous wide-field surveys, Ojha \etal\ (1996) 
derived the Galactic stellar density as a function of position and 
magnitude interval.  
From this, they derive scale lengths and heights of 
$h^{thin}_R = 2.3 \pm 0.6 ~$kpc; $h^{thin}_z = 0.26 \pm 0.05~$kpc 
and 
$h^{thick}_R = 3 \pm 1~$kpc; $h^{thick}_z = 0.76 \pm 0.05 ~$kpc  
for the thin and thick disk components of the Galaxy, respectively.
These results are in excellent agreement with those of 
Robin, Cr\'ez\'e, \& Mohan (1992) for the thin disk and 
Robin \etal\ (1996) who, on the basis of a maximum likelihood 
analysis of a variety of Galactic field samples from a variety of authors 
find 
$h^{thick}_R = 2.8 \pm 0.8 ~$kpc, and $h^{thick}_z = 0.76 \pm 0.05 ~$kpc. 
In a very recent study using a deep HST sample, 
Gould, Bahcall \& Flynn (1996) report an M-dwarf disk scale length of 
$h_R = 3.0 \pm 0.4~$kpc.  Their sample is most sensitive to stars 
far from the plane ($\sim$2 kpc).  
The work of Reid \& Majewski (1993) suggests that the thin and thick disk 
scale {\it heights\/} may be considerably larger than previously thought: 
they obtain $h^{thin}_z = 0.4~$kpc for less luminous (M-dwarf) tracers, 
and $h^{thick}_z = 1.5 \pm 0.1 ~$kpc. 
These authors also review the previous literature by others on the 
thick disk scale height, which typically lie in the lower 
range $0.7 < h^{thick}_z < 1.2~$kpc.  

In summary, recent evidence from a variety of studies 
drawing on different datasets suggests that the 
scale length of the old, thin disk is in the range  
$h^{thin}_R = 2.5 \pm 0.5~$kpc.  
If the scale height of the disk decreases inwards, but is modeled as 
remaining constant, estimates for $h_R$ may be biased artificially 
downwards, in which case the thin disk scale length 
may be more nearly $h^{thin}_R = 3.0 \pm 1~$kpc.  
The thick disk scale length $h^{thick}_R$ is probably similar to $h^{thin}_R$, 
though less well-constrained.  
Recent determinations of $h_R$ for the Galaxy are consistent 
with observations of external galaxies of the same type:  
the median scale length for Sb-Sc spirals derived from de~Jong's (1996) 
total sample of 86 face-on galaxies lies between 2.7 and 3.3~kpc 
(for $H_0 = 75 ~ {\rm km \, s^{-1} \, Mpc^{-1}}$), while the median 
scale length for Courteau's sample (1996) of over 300 spirals 
of varying inclination is 3.9 kpc for the same $H_0$ (1997, private 
communication).  
Galactic measurements of $h_R$ are relative measurements that 
generally scale with the assumed distance solar distance $\rnot$.  
As we will see, the dynamical importance of the disk is most sensitive 
to the ratio $\rnot/h_R$.  
Finally, most estimates for the local scale heights 
of the thin and thick components indicate that 
$h^{thin}_z = 0.30 \pm 0.05~$kpc and 
$h^{thick}_z = 1 \pm 0.25~$kpc, respectively.  

\subsection{Distance to the Galactic Center}

An excellent review of recent 
determinations of the distance $\rnot$ to the 
Galactic Center has been compiled by Reid (1993) and need not be 
repeated here.  
Reid concludes that the weight of the evidence 
suggested that $\rnot = 8.0 \pm 0.5~$kpc, with the error representing 
both systematic and statistical uncertainties.
We will assume this range for $\rnot$ here, which includes 
the current IAU standard value of 8.5~kpc as its upper limit.  

\subsection{The Rotation Curve of the Galaxy}

Whereas the rotation curves of external galaxies are relatively 
straight forward to measure, the rotation curve of the Galaxy has 
proven notoriously difficult to constrain 
(see reviews by Fich \& Tremaine 1991, 
Merrifield 1993; Schechter 1993).  
A review of all recent determinations of $\vnot$ is beyond 
the scope of this article, we concentrate on those works that 
illustrate the nature of the discrepancy remaining in this important 
Galactic measurement.

Interior to the solar position at $\rnot$, the tangent point  
(or terminal velocities) method 
produces accurate measurements (if the orbits are circular), but the 
velocities are measured relative to the local velocity $\vnot$ and 
distances are expressed in units of $\rnot$.  
In order to convert these data into an absolute rotation curve,  
assumptions must be made for the solar distance and local circular speed; 
the IAU standards of 
$\rnot = 8.5 \pm 1.1~$kpc and $\vnot = 220 \pm 20~\kms$, based 
on a review by Kerr \& Lynden-Bell (1986), have been  
typical choices in the past. 
Most recent estimates for both the local circular speed and distance 
to the Galactic center tend to be lower than the 1986 IAU standards.  
In the inner galaxy, estimates from the tangent point 
method yield a considerably lower value 
for $2  A  \rnot \equiv \vnot - \rnot \, (dv_c / dR)_{\rnot} 
\equiv - \rnot^2 \, (d \Omega / dR)_{\rnot}$, 
where $A$ is Oort's constant and $\Omega$ is the disk angular velocity. 
Comparing to previous analyses, this suggests one or more of the following: 
$\vnot$ is smaller; $\rnot$ is smaller; 
or the local rotation curve turns over less fast (\ie the Oort constant 
A is lower). 

Using the tangent point method on the 
neutral gas (\hi) component together with their new estimates for the 
solar distance and Oort constants, Rohlfs \etal\ (1986) 
derived a rotation curve that dropped from 
$200~\kms$ at 6 kpc to $170~\kms$ at 10~kpc, and flattened to $200~\kms$ 
again at large radii. The outer rotation curve was 
determined from \hii\ regions.   
Their best fit value of $\vnot = 184~ \kms$.   
They noted that their lower estimate 
for $\vnot$ actually produces a rotation curve for the Milky Way 
whose amplitude and slope more nearly matches that of external 
Sb-Sc spirals.  Larger values produced an atypical steeply-rising curve.  
 
Merrifield (1992) used an axisymmetric model of 
the neutral hydrogen layer to fit 
the Galactic \hi\ kinematics as a function of longitude and 
latitude, solving simultaneously for the form of 
the rotation curve (scaled to the solar distance and LSR speed) 
and the scale height of the gas as a function of radius.  
He was able to use this method out to $~2.5 \,\rnot$,   
with different choices for $\vnot$ and $\rnot$ leading to different 
slopes for the outer rotation curve.  Using the Oort constants of 
Kerr \& Lynden-Bell (1986), he finds that his data requires 
$\rnot = 7.9 \pm 0.8 \, $kpc and $\vnot = 210 \pm 25 \, \kms$. 
In order to be consistent 
with the rotation curve slopes of external galaxies of the 
same type, Merrifield (1992, 1993) concluded that $\vnot = 180 \pm 20~\kms$. 
(Larger $\vnot$ or smaller $\rnot$ will cause a sharp increase in 
the slope of the rotation curve, as noted by Rohlfs \etal\ 1986.)  
Merrifield thus compromised on $\vnot = 200 \pm 10~\kms$. 

In the outer Galaxy 
tangent points are not available, and the kinematics of 
standard candles are typically used, which are generally less reliable.  
These measurements tend to 
give higher values for $\vnot$; though this may in part be due 
to (slight) disk ellipticity.  

A recent determination from Cepheid work (Pont, Major \& Bruki 1994) 
measures the solar distance and Oort constants, and variation of 
the Cepheid kinematics with distance.  Together these 
allow a determination of Galactic rotation curve (which must be 
normalized with $\vnot$).  
Their solar distance of $\rnot = 8.09 \pm 0.3~$kpc 
falls nicely within other determinations, but their rather high 
Oort constant yields 
$2 A \rnot = 257 \pm 7 \, \kms$, which for a flat rotation curve 
would result in a very large $\vnot = 257 \, \kms$.  
Since their Cepheid rotation curve appears to fall in the region 
of the solar radius, however, the implied local circular speed $\vnot$ 
would probably be lower, by an amount that depends the gradient 
$(dv_c / dR)_{\rnot}$.  
Using their Oort constant $A$, but current IAU standards for 
$\rnot$ and $\vnot$, Pont \etal\ (1996) find that their outer rotation 
curve is flat at $200~\kms$.  They note 
that it is possible that their rotation curve is miscalibrated due to  
a systematic effect caused by 
metallicity gradients in the Cepheids with Galactocentric distance, 
and plan to design new observations to test this hypothesis.  
Such metallicity effects can lead to incorrect extinction 
corrections and thus incorrect Cepheid distances 
(Stothers 1988, Gould 1994, Beaulieu \etal\ 1996).

Schechter (1993) reports that carbon star kinematics 
in the outer Galaxy are consistent with a flat rotation curve of  
$227~ \kms$, with at least 10\% uncertainty from the 
uncertainty in the Local Standard of Rest (LSR) angular speed 
about the Galactic Center.  
This uncertainty may decrease soon as the VLA proper motions of 
Sgr A$^{*}$, which is believed to be stationary at the 
Galactic Center, improve with time.  
Backer (1996) reports an observed proper motion of Sgr A$^{*}$ 
in the longitudinal and latitudinal directions of 
$-6.55 \pm 0.34 ~ {\rm mas} \, {\rm yr}^{-1}$ and 
$-0.48 \pm 0.23 ~ {\rm mas} \, {\rm yr}^{-1}$, respectively.  
(Note that these are 2$\sigma$ errors.)  
This must be corrected to account for the motion of the Sun with 
respect to the LSR (Mihalas \& Binney 1981) tangent to the 
Galactic rotation of $+12 ~\kms$. 
For $\rnot = 8.0~$kpc, this leads to a local circular speed of 
$236 \pm 13 ~\kms$, considerably higher than estimates from 
the kinematics of \hi\ in the disk.

Kuijken \& Tremaine (1994) suggest that some of the discrepancy 
between $\vnot$ as determined from stellar and \hi\ tracers at 
different positions may be due to ellipticity of the galactic potential, 
and can be rectified if the disk isopotential curves 
are slightly elliptic to yield a true circular speed of about 
$\vnot = 200~\kms$.  
We conclude that the circular speed at the solar radius is probably 
lower than the standard IAU value and instead lies  
in the range $\vnot = 210 \pm 25 ~\kms$ which, assuming that 
$\rnot \approx 8~$kpc, encompasses the best 
values from the \hi\ analysis of $185 ~\kms$ on the low side and  
estimates based on the Sgr A$^{*}$ proper 
motion of $235 ~\kms$ on the high side. 

\subsection{A Word about the Bulge}

Sellwood \& Sanders (1988) were able to construct a maximum disk 
model for the Galaxy by assuming a disk scale length of 4 kpc 
and a very massive bulge of $(3-4) \times 10^{10} \, \msolar$.  
More recent estimates suggest that the bulge 
has a somewhat lower mass at $(1-2) \times 10^{10} \, \msolar$, 
and is likely, in fact, to be bar-like 
(see Zhao, Spergel \& Rich 1995 for a review), making a maximum 
disk solution less viable.  
On the other hand, the smaller Milky Way scale length indicated by 
more recent measurements operates in the other direction, 
as we shall see in \S 4.  
Note that the effect of the bulge is already taken into account 
by our definition of maximal through the spread in the 
fractional rotation support ($85 \pm 10\%$) provided by external 
maximal disks at $2.2h_R$ --- the more massive the bulges, 
the smaller the required disk support.  
The primary results of the next section thus do not depend on 
a specific bulge model.  
Given the uncertainty in the morphology and mass of the Galactic bulge, 
we do not draw any conclusions about bulge parameters, but 
do give one example a simple, illustrative model of the Galactic bulge 
to indicate roughly the magnitude of its effect on Milky Way kinematics. 

\section{A Maximum Disk in the Milky Way} 

The Milky Way is usually said to have a sub-maximal disk 
(Bahcall 1984; Kuijken \& Gilmore 1989b; van der Kruit 1989, 
Kuijken \& Gilmore 1991, Kuijken 1995).  
This statement is based on the observation that an 
exponential disk with a local surface mass density of 
$\Sigma_{\odot} \approx 50 \, \msolar pc^{-2}$ at an assumed 
$8 < \rnot\ < 8.5~$kpc, and a scale length $3.5 < h_R < 4.5~$kpc  
can provide only about half of the IAU-accepted local circular 
rotation speed of $\vnot = 220 \, \kms$. 
To provide all of the IAU estimate for $\vnot$, an 
exponential disk of this $h_R$ would need to have a local 
mass column of $\Sigma_{\odot} \approx 100 \, \msolar pc^{-2}$.

A more uniform definition of the ``maximum disk hypothesis,''  however, 
together with revised estimates for Galactic structure constants 
requires a reanalysis of this conclusion: 
\begin{itemize}
\item In external galaxies, the fraction of disk support 
	is usually computed at the point 
	where the disk rotation curve peaks at $\sim 2.2h_R$ (which 
	need not equal $\rnot$ in the Galaxy), or through 
	fits to the entire rotation curve.

\item Massive bulges and non-hollow halos require that a disk with 
	maximum mass actually provides less than 100\% of $v$(2.2$h_R$):  
	maximal disk fits in external Sb-Sc spirals typically give 
	$v^{disk}(2.2h_R)/v(2.2h_R) = 85 \pm 10\%$.  
	The Milky Way is known to have a massive bulge or bar, 
	though its precise dynamic contribution is ill-constrained. 

\item 	The total local column of the thin disk mass is probably 
	$\Sigma_{\odot} = 53 \pm 13 \, \msolar pc^{-2}$, but 
	the column within 1.1~kpc, 
	$\Sigma_{\rm 1.1~kpc} = 71 \pm 6 \, \msolar pc^{-2}$, is more robust.  
	A column of $50 \pm 10 \, \msolar pc^{-2}$ is probably 
	appropriate for the total {\it observed\/} then disk at $\rnot$.

\item Recent estimates for the scale lengths of the (old) thin and 
	thick disks remain somewhat uncertain, but 
	are lower than previous estimates;  
	$h_R = 3.0 \pm 1~\kpc$ is now an appropriate range.  Estimates for 
	the scale heights vary, but are generally bracketed by 
	$h^{thin}_z = 0.30 \pm 0.05~$kpc and $h^{thick}_z = 1 \pm 0.25~$kpc.
	A range of plausible disk $h_z$ must be considered because  
	rotation curve fitting in external galaxies is unable to place 
	constraints on the scale height of the disk {\it mass\/}.

\item New determinations for solar distance $\rnot$ are somewhat 
	lower than the 1986 IAU standard of 8.5~kpc, 
	with $8.0 \pm 0.5~$kpc providing a plausible range.

\item The latest determinations from both the \hi\ velocity field and 
	varying \hi\ scale height of the neutral gas 
	in the Milky Way indicate that the Galactic rotation curve 
	has a local circular speed that is 
	considerably lower than the IAU value of $220 \, \kms$, 
	but stellar tracers still tend to give higher values.  
	We choose $\vnot = 210 \pm 25 ~\kms$ as a plausible 
	range.  Note that use of the Tully-Fisher relation 
	(Pierce \& Tully 1988; Jacoby \etal\ 1992) indicates that 
	this velocity range corresponds to a total luminosity for 
	Milky Way of 
	$L_V \approx (2.1^{+1.0}_{-0.6}) \times 10^{10} \, L_{\odot, V}$, 
	consistent with observational estimates of 
	$(1.4 < L_V < 2.0) \times 10^{10} \, L_{\odot, V}$
	(van~der~Kruit 1986; Binney \& Tremaine 1987), with velocities 
	in the lower end of the range matching somewhat better.

\end{itemize}

Our goal is not to test any one choice for Galactic 
structure parameters, nor to argue that one set is preferable to another, 
but rather to ask whether the maximum disk hypothesis --- as it is 
applied in external galaxies --- is consistent with the newest 
constraints on the range of likely structural parameters for the Galaxy.
We will do this by computing the rotation support at 2.2$h_R$ 
supplied by various disk models as a function of the assumed 
ratio of solar distance to disk scale length $\rnot/h_R$, and then 
asking whether this rotation support lies within $85 \pm 10\%$ of 
the rotation speed at that distance. 
The parameter $\rnot/h_R$ is important because the maximum speed 
supplied by the disk is proportional to $\sqrt{M_{disk}}$,  
and the disk mass is normalized through a {\it local\/} 
measurement of the total column of disk mass through 
$M_{disk} = 2 \, \pi \, \Sigma_{\odot} \, h^2_R \, e^{\rnot/{h_R}}$.
We assume here that the rotation curve is approximately flat between 
$R = 2.2h_R$ and $R = \rnot$ so that 
$v(2.2h_R) \approx v(\rnot) \equiv \vnot$, and that any slight 
slope between these radii lies within the uncertainty in $\vnot$.

\subsection{The Old Definition and Old Constraints for a Galactic Maximal Disk}

Shown in Fig.~1 is the rotation support supplied by the disk at 
the solar position ($R = \rnot$) 
as a function of $\rnot/h_R$ for a variety of disk mass models and 
assumptions for the solar distance $\rnot$. 
Three mass models, none of which may be correct in 
detail, but which reasonably span the model space are considered: 
	(a) {\it Very thin, very light disk:\/}   
	the total mass column of the disk at 
	$\rnot$ is set equal to the minimum estimate of 
	$\Sigma_{\odot} = 40 \, \msolar pc^{-2}$ corresponding to a 
	lower limit on the   
	{\it directly observed\/} components alone, with 
	a small scale height of $h_z =~$0.25~kpc (solid lines).   
	(b) {\it Moderately thin, heavy disk:\/} 
	the disk mass is assumed to follow the distribution of thin 
	disk light with a scale height 
	$h_z =~$0.35~kpc, using an upper limit for the measured 
	local column of thin disk mass between 1.1~kpc  
	of $\Sigma_{\rm 1.1~kpc} = 65 \, \msolar pc^{-2}$ (dashed lines).   
	(c) {\it Thick, heavy disk:\/} 
	the disk mass is assumed to follow the distribution of thick 
	disk light with a scale height $h_z =~$1.0~kpc, and a 
	local column of disk mass between 
	1.1~kpc equal to 
	$\Sigma_{\rm 1.1~kpc} = 75 \, \msolar pc^{-2}$ (dotted lines).  
The rotation curve for each model was calculated in the manner described 
in \S2, taking into full account the thickness of the disk.   

For each mass model, two extreme choices for $\rnot$ are indicated:  
a small distance to the Galactic center of $\rnot = 7.5$~kpc, and  
the IAU standard value of $\rnot = 8.5$~kpc, which now 
appears to lie at the upper end of the probable range.   
At fixed $\rnot/h_R$, models with larger $\rnot$ must also have 
larger $h_R$, and thus larger total disk mass and larger rotation support 
($v \propto \sqrt{\rnot}$ for fixed $\rnot/h_R and \Sigma_{\odot}$).  
Models with higher local surface mass density have, naturally, 
a higher total mass at fixed $\rnot/h_R$, and thus more rotation support 
($v \propto \sqrt{\Sigma_{\odot}}$ for fixed $\rnot/h_R and \rnot$).
Moving along a given model line toward increasing $\rnot/h_R$ is 
equivalent to decreasing the disk scale length $h_R$ for that model.
The disk support increases for smaller $h_R$ at fixed $\rnot$ 
because the constant {\it local\/} normalization for the disk 
column then occurs at a larger number of disk scale lengths; this 
has an exponential effect on total disk mass that more than offsets 
the decrease in $M_{disk}$ with decreasing $h_R$ 
($v \propto \sqrt{h_R \, e^{1/{h_R}}}$ for fixed $\rnot and \Sigma_{\odot}$).
Disk models that are constrained by a local 
column within 1.1~kpc are more massive than those constrained to have 
the same total column, and thus produce larger disk support at the 
same $\rnot/h_R$.

Fig.~1 shows how the maximum disk hypothesis 
has been tested previously in the Milky Way: the rotation support 
of a particular Galactic disk model at $R = \rnot$ 
is compared to the IAU standard value of $\vnot = 220 ~ \kms$.
The popular models of Kuijken \& Gilmore (1991, KG) and 
Bahcall \& Soneira (1980, BS, with $\Sigma_{\odot}$ normalization from 
Bahcall 1984) are indicated in Fig.~1;  both 
clearly fall far short of providing all of IAU standard circular speed.
Indeed, this is the primary basis for previous statements that 
the Galactic disk is approximately ``half-maximal.''

\subsection{Application of the New Definition}

Fig.~2 indicates the situation changes 
when using the extragalactic 
definition of maximal disk, which makes the comparison at $R = $2.2~$h_R$, 
(not $R = \rnot$), and more recent Galactic constraints.  
The disk rotation support is now computed at its  
maximum at (\ie $R = 2.2 h_R$), 
and plotted in Fig.~2 for comparison with the 
speed that the disk must have in order to be considered maximal by the 
definition used in extragalactic studies (see \S2). 
The ordinate bounds of the box in Fig.~2 
indicate the upper and lower limits of the current observational 
constraints on the local circular speed $\vnot = 210 \pm 25 ~\kms$ 
multiplied by the fraction $85 \pm 10\%$ of typical maximum disk support 
in external Sb-Sc galaxies at 2.2~$h_R$.
The abscissa bounds of the box indicate the constraints on 
$\rnot/h_R$ (see Table 1).  
Considering the observational errors, $ 2 < \rnot/h_R < 4.8$, but note that 
the best fit value of all recent determinations cluster in 
the range $2.7 < \rnot/h_R < 3.5$.  
Models falling inside the 
box are simultaneously ``maximal'' and 
satisfy observational constraints on $\rnot/h_R$; those above the box 
are too massive and those below are sub-maximal. 
The dynamical constraints actually rule out certain combinations 
of $R/h_R$ for the Milky Way that are allowed by the structural constraints 
alone.  In particular, $\rnot/h_R > 4.3$ is excluded since then 
all disk models would be so massive as to exceed estimates for the 
local rotation speed, even without the addition of mass 
components such as the bulge and dark halo.   

Models constrained to the ``maximal disk box'' have total disk masses 
that vary with $\rnot/h_R$ and lie in the range:  
$3.1 \times 10^{10} < M < 6.7 \times 10^{10} \msolar$ for the very thin, 
very light disk, 
$4.4 \times 10^{10} < M < 8.0 \times 10^{10} \msolar$ for the moderately thin, 
heavy disk, and
$7.3 \times 10^{10} < M < 1.1 \times 10^{11} \msolar$ for the thick, 
heavy disk.  
Using observational determinations of the disk V band luminosity 
of the Galactic disk of $\sim 1.5 \times 10^{10} L_{\odot, V}$
(van~der~Kruit 1986; Binney \& Tremaine 1987),   
the full range of the implied mass-to-light ratios for these maximal disk 
models is thus $2 \lsim M/L_V \lsim 7$, which compares quite 
well with typical values of $2 \lsim M/L_V \lsim 6$ 
found for maximum disks of external Sb-Sc spirals 
(Begeman 1987; Broeils 1992; Broeils \& Courteau 1997).

{\it All\/} the disk models presented in Fig.~2 could be maximal 
if $\rnot/h_R \geq 3$ (\ie the scale length of the disk mass is 
$h_R \ltorder 2.5$~kpc), which is within the observational constraints.     
The heavier disk models, those with larger disk scale height, 
and those assuming a larger solar distance, can be maximal 
for larger values of $h_R$. 
The KG model still falls below the Galactic maximum disk box; but 
also just to the left, indicating that their assumed $\rnot/h_R$  
may no longer be realistic.  
If the KG model were changed only by reducing the assumed disk 
scale length from 4.5~kpc to 3~kpc, it could be a maximum disk model.  
The BS model clearly lies within the maximum disk box; changing 
their assumed $h_R$ from 3.5~kpc to 3~kpc would place close to the 
middle of the velocity limits.  

\subsection{Addition of an Illustrative Bulge Model}

Rather than model the bulge directly, the analysis of the previous 
section used as a constraint 
the range of rotation support supplied by the 
maximum disks of external Sb-Sc spirals, 
namely $85 \pm 10\%$ of the total $v(2.2 h_R)$. 
This range encompasses the range of bulge masses typically seen in 
spirals similar in type to the Galaxy, and thus represents a reasonable   
range for the Milky Way.  
If the mass and profile of the Galactic bulge were known precisely, 
tighter constraints could be placed, but the structure of the bulge 
is a matter of some debate (\eg Zhao, Spergel \& Rich 1995), 
and so we have chosen not to model it directly.  
Nevertheless, in order to assess (roughly)   
the dynamical influence of the bulge on the possibility 
of a Galactic maximal disk we now turn to a simple, empirical model for 
the bulge.  

A detailed mass model of the Galaxy would likely include 
separate components for the Galactic Center, inner bulge, bar, thick disk, 
and metal-weak halo.  Since the combined uncertainty in the structure of 
these components is large and controversial, we prefer to use 
the spherical spheroid model of de~Vaucouleurs \& Pence (1978), which  
has a simple analytic form and is meant to account for all the non-disk 
(thin disk) light in the Galaxy.  
For illustrative purposes, we therefore consider a spheroidal mass 
component with an $R^{1/4}$ profile, 
and an effective radius and total mass scaled to give 
$R_e = 2.67~$kpc and $M_{bulge} = 1.5 \times 10^{10} \, \msolar$, 
respectively, for an assumed solar distance of $\rnot = 8~$kpc.  
Models with different choices for $\rnot$ are scaled appropriately. 

In Fig.~3, the rotation support of this scaled bulge model has been added 
to that of each of the disk models to 
arrive at an estimate of total rotation supplied by the mass associated 
with the luminous components of the Milky Way at $R = 2.2 \, h_R$.  
The ordinate bounds of the maximal disk box have now been changed 
to reflect 100\% of the observational limits on the Galactic rotation 
curve.  Again we see that {\it all\/} disk models can be considered 
maximal for some (allowable) range of Galactic parameters, and 
the models nicely bracket 
the most probable range of $\rnot/h_R$ indicated by the smaller inset box.  
The heavy, thin disk model, for example, with $h_z = 0.35~$kpc and 
a local mass column between 1.1~kpc 
of $\Sigma_{\rm 1.1~kpc} = 65 \, \msolar pc^{-2}$ 
(or total column $\Sigma_{\solar} = 68 \, \msolar pc^{-2}$), 
could be maximal over the full range of the median $\rnot/h_R$ 
from recent Galactic observations, in the sense that the 
dark halo supplies {\it none\/} of the rotation support at $R = 2.2 \, h_R$.  

\section{Conclusions and Implications}

Recent determinations for both the Galactic disk scale length, and 
the magnitude and slope of the rotation curve at the solar position 
are smaller than older values and more comparable to those 
determined for external Sb-Sc spirals, indicating that the Milky Way 
disk may provide more dynamical  
support in the inner Galaxy than previously thought. 
The maximum disk hypothesis applied to external Sb-Sc galaxies yields 
a disk mass that provides $85 \pm 10\%$ of the total 
rotation support at $2.2 h_R$, where the disk support peaks. 
{\it Combining this definition of maximal with new observational constraints 
on Galactic structure parameters leads to the conclusion that 
it is plausible --- and perhaps even likely --- 
that the Galactic disk is maximal, in the sense that the dark halo 
provides negligible rotation support inside two disk scale lengths.  
In particular, given the new Galactic constraints, the local disk 
column can be substantially less than the oft-quoted required 
$\Sigma_{\odot} \approx 100 \, \msolar pc^{-2}$ and still be maximal.\/}  
The maximal disk models presented here have mass-to-light ratios  
that lie precisely in the range determined for maximal disks in 
external Sb-Sc spirals. 

On the other hand, if the disk scale length for the Milky Way could be shown 
definitively to be above 4~kpc, most thin maximal disk models 
would be ruled out.  
Obtaining tight constraints on the Galactic disk scale length, 
solar distance, and rotation curve is  
thus as important as constraining the local surface mass density 
in assessing the dynamical dominance of the disk in the inner Galaxy. 
 
A Milky Way maximal disk would alter estimates for Galactic dark 
halo parameters, since the two combined (with the addition 
of the bar/bulge) give rise to the Galaxy's rotation curve.  
Any astronomical conclusion, therefore, based on assumed potentials  
for the disk or dark halo of the Galaxy may need to be 
revised in light of the possibility that the Galactic disk is 
maximal.  In particular, theoretical estimates for the 
microlensing optical depth along both the Galactic Center and Magellanic 
Clouds lines of sight may be in need of revision: a maximal 
disk would, in general, result in a higher optical depth toward the bulge 
due to disk lenses, and --- since the halo would then be a less dominant 
component of the inner Galaxy --- a lower optical depth along 
both sight lines due to halo lenses. 


\acknowledgments 

It is a pleasure to thank Tjeerd van~Albada, Andy Gould, and Konrad Kuijken for 
useful comments and a critical reading of the manuscript.  
This work was partially funded by the National Science Foundation 
(AST~92-15485).



\clearpage 


\begin{deluxetable}{lccccc}
\tablenum{1}
\tablecaption{
Recent Determinations of Galactic Scale Parameters with assumed $\rnot$
}
\tablehead{
\colhead{Reference} & \colhead{Disk Type} 
& \colhead{$h_R$} & \colhead{$h_z$}
& \colhead{$\rnot$} & \colhead{$\rnot/h_R$} \nl
 & & \colhead{(kpc)} & \colhead{(kpc)} & \colhead{(kpc)} & 
}
\startdata
Kent, Dame \& Fazio 1991 
   & 2.4 micron & $3.0 \pm 0.5$ & 0.247\tablenotemark{a} & 8.0 & 2.7 \nl 
Robin, Cr\'ez\'e, \& Mohan 1992  
   & thin & $2.5 \pm 0.3$ & constant & 8.5 & 3.4 \nl
Fux \& Martinet 1994 
   & thin & $2.5^{+0.8}_{-0.6}$ & constant & 8.5 & 3.4 \nl
Fux \& Martinet 1994 
   & thin & $3.1$ & flaring\tablenotemark{b} & 8.5 & 2.7 \nl
Ojha et al. 1996 
   & thin & $2.3 \pm 0.6$ & $0.26 \pm 0.05$ & 8.1 & 3.5 \nl 
Ojha et al. 1996 
   & thick & $3 \pm 1$ & $0.76 \pm 0.05$ & 8.1 & 2.7 \nl
Robin et al. 1996 
   & thick & $2.8 \pm 0.8$ & $0.76 \pm 0.05$ & 8.5 & 3.0 \nl
Gould, Bahcall \& Flynn 1996 
   & M-dwarf & $3.0 \pm 0.4$ & 0.45\tablenotemark{c} & 8.0 & 2.7 \nl
Reid \& Majewski 1993
   & M-dwarf & --- & $0.4$ & --- & --- \nl
Reid \& Majewski 1993
   & thick & --- & $1.5 \pm 0.1$ & --- & --- \nl

\enddata
\tablenotetext{a}{at $\rnot$, and decreasing inwards to 0.165 kpc at the 
Galactic center}
\tablenotetext{b}{assumed to rise with $R$ at rate of 30~pc/kpc}
\tablenotetext{c}{best fit with single component; two components provide  
better fit}

\end{deluxetable}

\clearpage
  

\vskip 1in
\begin{figure}
\hsize 6.5in
\epsfxsize=\hsize\epsffile{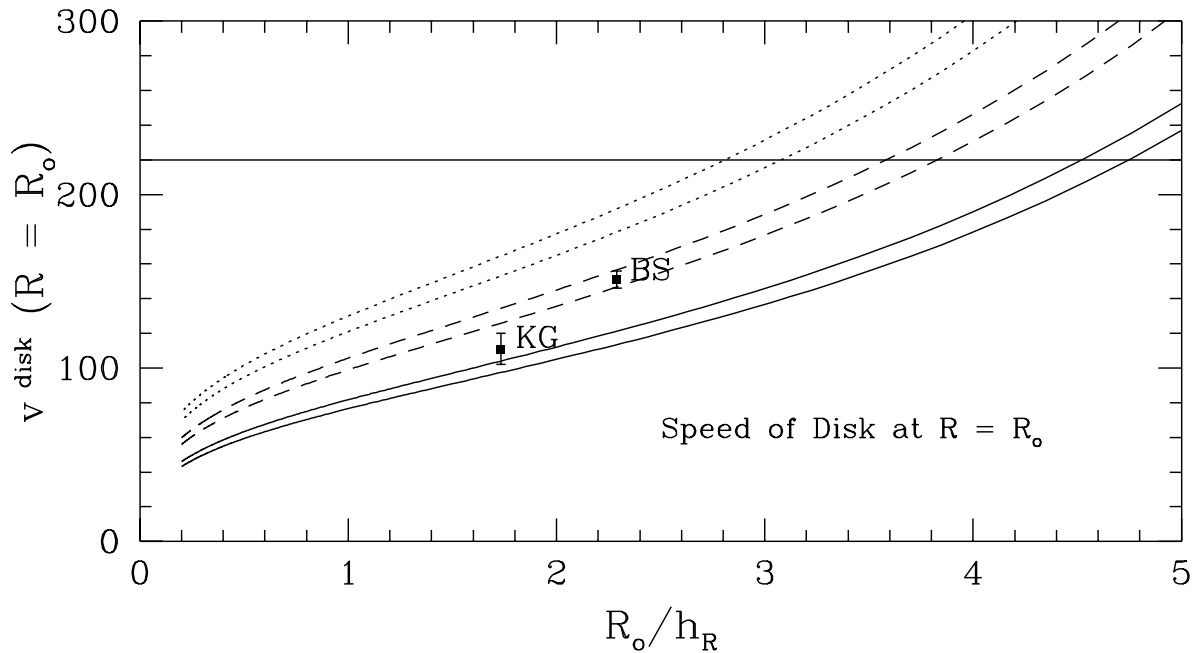}
\vskip -2in
\caption{
The rotation support at the solar position supplied by the disk alone 
as a function of $\rnot/h_R$ for three disk mass models that 
are explained more fully in the text: 
(a) very thin, very light disk (solid lines), 
(b) moderately thin, heavy disk (dashed lines), and
(c) thick, heavy disk (dotted lines).  
For each mass model, 
two assumptions for the solar distance are indicated: the lower  
lines in each pair assume $\rnot =~$7.5~kpc; the upper 
lines take $\rnot =~$8.5~kpc.  
Any point along these curves represents a particular choice for the 
disk mass model and the structure constants $\rnot$ and $h_R$.
The Galactic disk models of Kuijken \& Gilmore (1991, KG) and 
Bahcall \& Soneira (1980, BS) are shown.  The horizontal line 
indicates the current IAU standard for the local circular 
velocity $\vnot = 220~ \kms$; new measurements indicate that this may 
be too high. 
}
\end{figure}


\vskip 1in
\begin{figure}
\hsize 6.5in
\epsfxsize=\hsize\epsffile{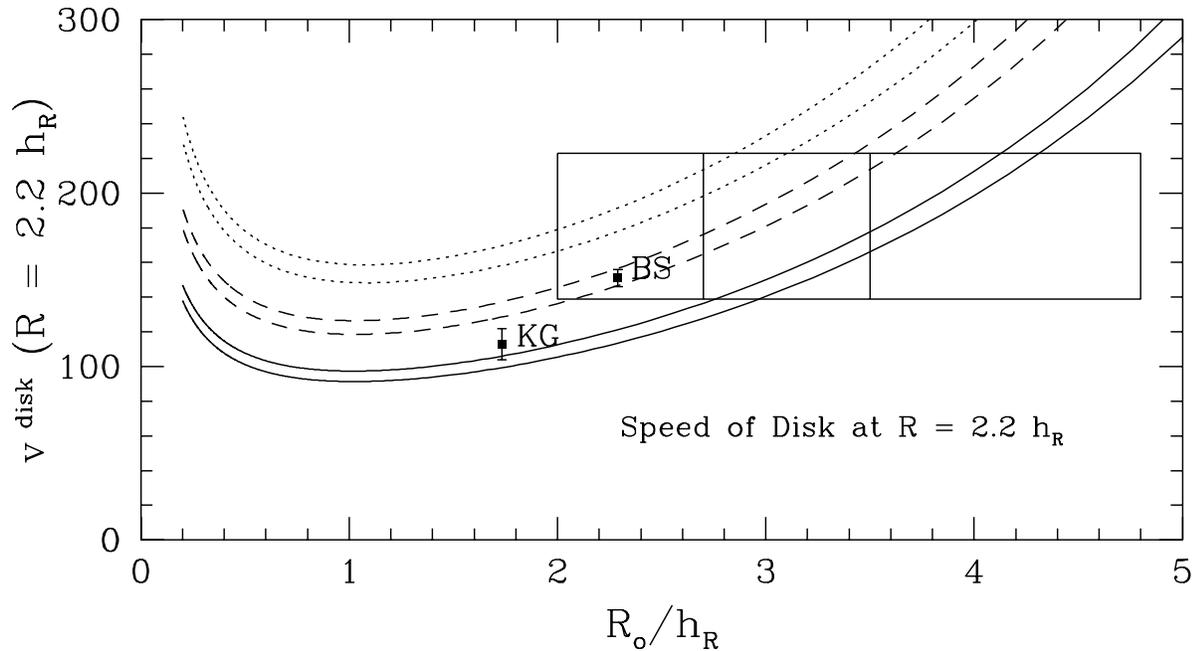}
\vskip -2in
\caption{
The rotation support at $R = 2.2 h_R$ supplied by the disk alone 
(at the position of maximum disk support) 
as a function of $\rnot/h_R$ for the same three models displayed in Fig.~1.
The long narrow box encloses the 
maximum disk constraint for the Galaxy;   
disk models falling inside the box can be considered to be maximal 
and satisfying current constraints on $\rnot/h_R$ 
(see text).  The smaller box delineates the range of best fit values 
for $\rnot/h_R$ from Table 1. 
}
\end{figure}


\vskip 1in
\begin{figure}
\hsize 6.5in
\epsfxsize=\hsize\epsffile{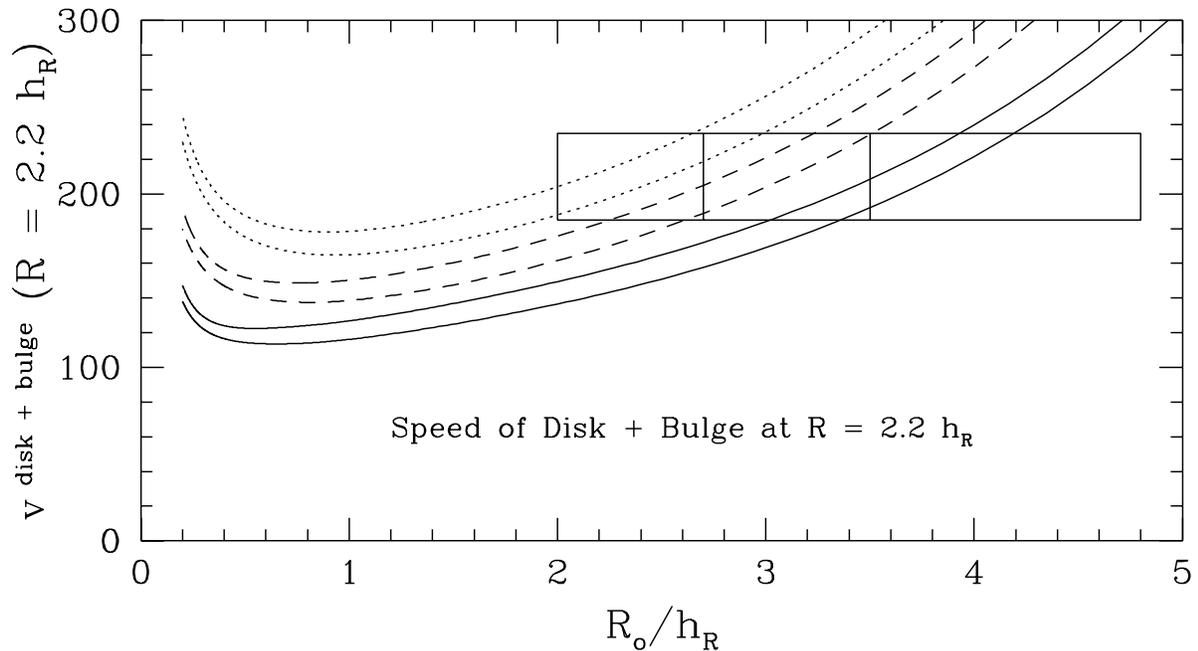}
\vskip -2in
\caption{
The rotation support at $R = 2.2 h_R$ supplied by the disk and bulge 
combined as a function of $\rnot/h_R$ for the same three models displayed 
in Fig.~1.  The simple illustrative model for the bulge is explained 
more fully in text, and is scaled to have the same observed luminosity 
for all $\rnot$, giving a range of 
$1.2 \times 10^{10} < M_{bulge} < 1.8 \times 10^{10} \, \msolar$.   
The long narrow box encloses a strict ``maximum luminous mass''
constraint for the Galaxy:    
mass models that fall inside the box satisfy constraints on 
$\rnot/h_R$, and together the disk and bulge provide {\it all\/} the rotation 
support at $R = 2.2 h_R$.  
The smaller box delineates the range of best fit values 
for $\rnot/h_R$ from Table 1. 
}
\end{figure}
\end{document}